\documentclass[11pt,a4paper]{article}

\usepackage[T1]{fontenc}
\usepackage{lmodern}
\usepackage{graphicx}
\usepackage{booktabs}
\usepackage{array}
\usepackage{amsmath,amssymb,bm}
\usepackage[margin=1in]{geometry}
\usepackage{xurl}
\usepackage[hidelinks]{hyperref}

\title{Precision-Era Reassessment of Modified Quark--Lepton Complementarity: NuFIT 6.1, Phase-Convention Covariance and Lepton-Number Violation}

\author{%
Gazal Sharma$^{a,b,*}$ and Gaurav Katoch$^{c,d}$\\[0.8em]
\footnotesize $^{a}$Bahra Research Innovation \& Knowledge Cluster, Department of Physics,\\
\footnotesize Rayat Bahra University, Mohali, Punjab, India\\
\footnotesize $^{b}$Centre of Research Impact and Outcome, Chitkara University, Rajpura 140417, Punjab, India\\
\footnotesize $^{c}$Department of Physics, Rayat Bahra University, Mohali, Punjab, India\\
\footnotesize $^{d}$Department of Physics, Sharda School of Engineering \& Science,\\
\footnotesize Sharda University, Greater Noida, India\\[0.5em]
\footnotesize $^{*}$Corresponding author: \href{mailto:gazzal.sharma555@gmail.com}{gazzal.sharma555@gmail.com}\\
\footnotesize \href{mailto:gauravkatoch1988@gmail.com}{gauravkatoch1988@gmail.com}
}

\date{}

\begin{document}
\maketitle

\begin{abstract}
Quark--lepton complementarity (QLC) provides a phenomenological framework for comparing quark and lepton flavour structures. We reassess a modified QLC correlation-matrix construction using the 2026 Particle Data Group quark-mixing inputs and the NuFIT 6.1 oscillation likelihoods. The historical prediction $\sin^2\theta_{23}=0.4235$ is strongly disfavoured, with $\Delta\chi^2=17.99$ for normal ordering and 20.43 for inverted ordering, whereas later ordering-dependent estimates remain compatible with present preferred regions. Reconstructed ensembles of the complex correlation matrix show that the first row is comparatively stable, while about 99\% of the squared mean-texture evolution occurs in the lower two rows; roughly 72\% of the present ensemble remains closer to the tribimaximal than to the bimaximal reference texture.

We further formulate the lepton sector in the symmetric Schechter--Valle parametrization. The oscillation phase appears as the invariant combination $\delta=\phi_{13}-\phi_{12}-\phi_{23}$, while independent phase directions remain relevant to lepton-number-violating (LNV) amplitudes. We prove that, for unrestricted diagonal quark--lepton mismatch phases, changing from the standard PDG convention to the symmetric parametrization leaves the full ensemble of $|V_{c,ij}|$ invariant: additional left phases are absorbed into the mismatch matrix and right phases only rephase columns. Consequently, the QLC magnitude texture does not constrain the independent Majorana/LNV phases. The historical $S_1$ and $S_2$ limits are also found to lie close to their phase-agnostic kinematic ceilings.
\end{abstract}


\section{Introduction}
\label{sec:introduction}

The origin of fermion flavour mixing remains an open problem. In the Standard Model, quark mixing is described by the Cabibbo--Kobayashi--Maskawa (CKM) matrix, while neutrino oscillations probe the Pontecorvo--Maki--Nakagawa--Sakata (PMNS) matrix \cite{Maki:1962mu,Cabibbo:1963yz,Kobayashi:1973fv}. Their measured structures are markedly different: the CKM matrix is close to the identity and strongly hierarchical, whereas the PMNS matrix contains two large mixing angles and a smaller but non-zero reactor angle. The possibility that the two sectors nevertheless reflect a common flavour structure has motivated quark--lepton complementarity (QLC) and related unification-inspired mixing schemes.

The QLC idea and its possible theoretical origins were developed in a series of early studies. Raidal discussed relations between quark and lepton mixing angles in grand-unified settings \cite{Raidal:2004iw}; Minakata and Smirnov formulated general conditions for QLC and examined its phenomenological consequences \cite{Minakata:2004xt}; possible gauge-theoretic and unified realizations were investigated by Frampton and Mohapatra \cite{Frampton:2004vw} and by Antusch, King and Mohapatra \cite{Antusch:2005ca}. More general non-trivial correlations between the CKM and MNS/PMNS matrices were also considered by Xing \cite{Xing:2005ur}. Subsequent analyses explored next-to-leading-order QLC relations \cite{Harada:2013wxa}, correlation-matrix predictions for the reactor angle \cite{Chauhan:2006im,Picariello:2007kg}, and departures from idealized tribimaximal (TBM) and bimaximal (BM) textures \cite{Shimizu:2010qy}. We emphasize these antecedents because the present work is a reassessment of an established framework, not a claim to introduce QLC itself.

The formulation examined here was studied in detail by Sharma and Chauhan in 2016 \cite{Sharma:2016epj}. It replaces simple angle-sum rules by a non-trivial correlation matrix,
\begin{equation}
 V_c=U_{\rm CKM}\,\Psi\,U_{\rm PMNS},
 \qquad
 \Psi={\rm diag}(e^{i\psi_1},e^{i\psi_2},e^{i\psi_3}),
 \label{eq:qlc-master}
\end{equation}
where $\Psi$ accounts for possible phase mismatch between the quark and lepton sectors. In the present manuscript we use the shorthand \emph{modified QLC} (MQLC) for this non-trivial correlation-matrix implementation, distinguishing it from the simplest angle-sum version of QLC. This terminology does not imply a new modification introduced here; the relation and its motivation are inherited from the cited literature and the previous analysis.

Using the data then available, ref.~\cite{Sharma:2016epj} reconstructed the probability distributions of the nine elements of $V_c$, found the preferred texture to be closer to TBM than BM, and obtained the narrow atmospheric-angle estimate
\begin{equation}
 \sin^2\theta_{23}=0.4235^{+0.0032}_{-0.0043}.
 \label{eq:oldtheta23}
\end{equation}
The same analysis reported $|J_{\rm CP}|<0.0315$ together with bounds on two Majorana-sensitive invariants. A later hierarchy-dependent implementation gave central atmospheric-angle values of $44.24^\circ$ for normal hierarchy and $47.16^\circ$ for inverted hierarchy \cite{Sharma:2018hierarchy}. Related attempts to connect quark and lepton mixing continue to appear in the contemporary literature, including seesaw-based constructions that derive CKM--PMNS relations from restricted mass structures \cite{Albergaria:2024}. These developments make it useful to distinguish the broad idea of quark--lepton correlation from individual numerical predictions obtained at a particular stage of experimental precision.

The phase structure deserves separate attention when neutrinos are Majorana particles. The standard PDG convention is optimized for oscillation phenomenology, whereas the original symmetric parametrization of Schechter and Valle associates a phase with each elementary lepton rotation and makes the separation between oscillation CP violation and lepton-number-violating phases more transparent \cite{SchechterValle1980,RodejohannValle2011}. This distinction is also natural in seesaw-based descriptions of neutrino mass and in modern symmetry approaches to flavour \cite{DingValle2025,AvilaEtAl2026}. We therefore use the symmetric representation below to determine which features of the MQLC correlation texture are genuinely convention robust and which phase information is not fixed by the present construction.

The experimental situation has changed substantially since 2016. NuFIT 6.1 incorporates solar, reactor, accelerator and atmospheric information available through November 2025 and supplies numerical one-, two- and three-dimensional $\Delta\chi^2$ projections of the global three-flavour fit \cite{Esteban:2024eli,NuFIT61}. In particular, the atmospheric angle and the Dirac CP phase remain correlated and ordering dependent. The quark-sector inputs are also more precise in the 2026 Review of Particle Physics \cite{PDG2026}. The resulting data allow two questions to be addressed separately: whether earlier MQLC predictions survive as genuine out-of-sample tests, and how much the inferred correlation texture itself has moved as the PMNS parameter space has tightened.

That separation is essential. Reconstructing $V_c$ from the present PMNS matrix and then immediately inverting the same relation would be algebraically circular. We therefore do not use the present reconstruction to claim a new prediction of $\theta_{23}$. Instead, the previously published atmospheric-angle values are evaluated directly against the NuFIT 6.1 likelihood profiles. The present-day $V_c$ ensemble is used only to study the stability, ordering dependence and reference-texture proximity of the MQLC correlation matrix. We also retain the complete complex matrix for every Monte Carlo realization rather than treating an element-by-element average of moduli as a physical unitary matrix.

The paper is organized as follows. Section~\ref{sec:framework} summarizes the modified QLC framework, introduces the symmetric lepton-mixing parametrization and establishes the phase-convention covariance of the magnitude texture. Section~\ref{sec:method} describes the historical benchmark, current inputs and projection-weighted sampling procedure. Section~\ref{sec:historical-test} presents the retrospective tests of the earlier atmospheric-angle and CP results, including a reappraisal of the Majorana-sensitive invariants. Section~\ref{sec:vc-results} gives the precision-era correlation texture and its ordering dependence. Section~\ref{sec:tbm-robust} discusses the TBM/BM comparison and robustness checks, followed by the limitations and conclusions in sections~\ref{sec:discussion} and \ref{sec:conclusions}.

\section{Modified QLC framework}
\label{sec:framework}

\subsection{Mixing matrices and the correlation relation}

The observable quark and lepton mixing matrices originate from mismatches between the unitary rotations that diagonalize the corresponding fermion mass matrices. Writing
\begin{equation}
 U_{\rm CKM}=U_u^\dagger U_d,
 \qquad
 U_{\rm PMNS}=U_\ell^\dagger U_\nu,
 \label{eq:mixingdefinitions}
\end{equation}
relations between quark and lepton Yukawa matrices in unified constructions can induce correlations between the observable mixing matrices. The MQLC relation used here is eq.~\eqref{eq:qlc-master}. We treat it phenomenologically and do not assume a specific ultraviolet completion.

We work throughout in the standard three-flavour PMNS convention,
\begin{equation}
U_{\rm PMNS}=U_D P_M,
\label{eq:pmnsfactor}
\end{equation}
with
\begin{equation}
U_D=\begin{pmatrix}
 c_{12}c_{13} & s_{12}c_{13} & s_{13}e^{-i\delta}\\
 -s_{12}c_{23}-c_{12}s_{23}s_{13}e^{i\delta} &
 c_{12}c_{23}-s_{12}s_{23}s_{13}e^{i\delta} & s_{23}c_{13}\\
 s_{12}s_{23}-c_{12}c_{23}s_{13}e^{i\delta} &
 -c_{12}s_{23}-s_{12}c_{23}s_{13}e^{i\delta} & c_{23}c_{13}
\end{pmatrix},
\label{eq:pmns}
\end{equation}
where $s_{ij}\equiv\sin\theta_{ij}$ and $c_{ij}\equiv\cos\theta_{ij}$. For Majorana neutrinos,
\begin{equation}
P_M=\operatorname{diag}(e^{i\alpha_1},e^{i\alpha_2},1),
\label{eq:majorana}
\end{equation}
while $P_M=\mathbb{1}$ for Dirac neutrinos.

The Majorana phases are irrelevant for the observables studied here. Inserting eq.~\eqref{eq:pmnsfactor} in eq.~\eqref{eq:qlc-master} gives
\begin{equation}
 V_c=\left(U_{\rm CKM}\Psi U_D\right)P_M.
 \label{eq:majrephase}
\end{equation}
The multiplication by $P_M$ rephases complete columns of $V_c$, so that every $|V_{c,ij}|$ is independent of $\alpha_{1,2}$. We therefore do not sample Majorana phases when reconstructing the magnitude texture. The three $\psi_i$ are sampled uniformly over $[0,2\pi)$, following the historical analysis. Only relative combinations affect $|V_c|$, so retaining all three merely preserves the original convention.

\subsection{Symmetric parametrization, LNV amplitudes and phase covariance}
\label{sec:symmetric}

The standard PDG form is convenient for oscillation analyses, but it does not make the separation between oscillation CP violation and lepton-number-violating (LNV) phases manifest. An equivalent symmetric parametrization, originally introduced by Schechter and Valle and later discussed in detail by Rodejohann and Valle \cite{SchechterValle1980,RodejohannValle2011}, writes the three-neutrino mixing matrix as
\begin{equation}
 K=\omega_{23}(\theta_{23},\phi_{23})\,
 \omega_{13}(\theta_{13},\phi_{13})\,
 \omega_{12}(\theta_{12},\phi_{12}),
 \label{eq:symmetricK}
\end{equation}
where a phase is associated with each elementary rotation. The single phase combination measured by ordinary three-flavour oscillations is
\begin{equation}
 \delta=\phi_{13}-\phi_{12}-\phi_{23},
 \label{eq:symmetricdelta}
\end{equation}
for the rotation convention adopted here. The corresponding Jarlskog invariant is
\begin{equation}
 J_{\rm CP}=\frac{1}{8}\sin2\theta_{12}\sin2\theta_{23}
 \sin2\theta_{13}\cos\theta_{13}
 \sin(\phi_{13}-\phi_{12}-\phi_{23}),
 \label{eq:symmetricJ}
\end{equation}
which is identical to the standard result after using eq.~\eqref{eq:symmetricdelta}.

For this convention the symmetric matrix is related exactly to the Dirac part of the PDG matrix by
\begin{equation}
 K=D_L\,U_D(\delta)\,D_R,
 \label{eq:KDUD}
\end{equation}
with
\begin{align}
 D_L&=\operatorname{diag}\!\left(1,e^{i\phi_{12}},e^{i(\phi_{12}+\phi_{23})}\right),\\
 D_R&=\operatorname{diag}\!\left(1,e^{-i\phi_{12}},e^{-i(\phi_{12}+\phi_{23})}\right).
 \label{eq:DLDR}
\end{align}
Inserting eq.~\eqref{eq:KDUD} into the MQLC relation gives
\begin{align}
 V_c^{\rm(sym)}
 &=U_{\rm CKM}\,\Psi\,D_L\,U_D(\delta)\,D_R \nonumber\\
 &=U_{\rm CKM}\,\Psi'\,U_D(\delta)\,D_R,
 \qquad \Psi'\equiv\Psi D_L.
 \label{eq:Vcsym}
\end{align}
Because $\Psi$ and $D_L$ are diagonal, the transformation merely shifts the mismatch phases,
\begin{equation}
 (\psi_1,\psi_2,\psi_3)\rightarrow
 \left(\psi_1,\psi_2+\phi_{12},\psi_3+\phi_{12}+\phi_{23}\right).
 \label{eq:psishift}
\end{equation}
In the present analysis all three $\psi_i$ are independently sampled over the full interval $[0,2\pi)$. The transformed matrix $\Psi'$ therefore has exactly the same probability measure as $\Psi$. The factor $D_R$ only rephases complete columns and cannot change any matrix modulus. It follows that
\begin{equation}
 p_{\rm sym}\!\left(|V_{c,ij}|\right)=p_{\rm PDG}\!\left(|V_{c,ij}|\right)
 \qquad\text{for every }i,j.
 \label{eq:moduluscovariance}
\end{equation}
Equation~\eqref{eq:moduluscovariance} is stronger than a numerical robustness check: it is an exact phase-convention covariance of the magnitude texture under the assumptions of the present MQLC construction. Consequently, the nine $|V_{c,ij}|$ distributions, their overlap coefficients, the TBM/BM comparison and the mean-texture drift are unchanged by replacing the standard PDG representation with the symmetric Schechter--Valle form. The complex phase content of $V_c$, by contrast, is not fixed unless an ultraviolet theory also specifies the mismatch phases.

The physical advantage of the symmetric parametrization is most visible when oscillation and LNV amplitudes are written side by side. Ordinary oscillation probabilities are built from combinations such as
\begin{equation}
 U_{\alpha i}U_{\beta i}^{*}U_{\alpha j}^{*}U_{\beta j},
 \label{eq:oscphasecancel}
\end{equation}
so a phase attached to an entire mass-eigenstate column cancels. By contrast, a light-Majorana-neutrino contribution to an LNV amplitude has schematically the phase structure
\begin{equation}
 \mathcal A_{\alpha\beta}^{\rm LNV}\propto
 \sum_i m_i\,U_{\alpha i}U_{\beta i}\,\mathcal F_i,
 \label{eq:lnvamplitude}
\end{equation}
where $\mathcal F_i$ denotes process-dependent kinematic and hadronic/nuclear factors. The absence of complex conjugation means that the Majorana phases remain physical in such amplitudes. For neutrinoless double-beta decay with light-neutrino exchange, the relevant effective mass is
\begin{equation}
 m_{\beta\beta}=\left|
 c_{12}^{2}c_{13}^{2}m_1+s_{12}^{2}c_{13}^{2}m_2e^{-2i\phi_{12}}
 +s_{13}^{2}m_3e^{-2i\phi_{13}}
 \right|,
 \label{eq:mbb-symmetric}
\end{equation}
so the two LNV phase directions appear directly and independently of $\phi_{23}$. This is the principal conceptual advantage of the symmetric description for LNV phenomenology: the phase combination that governs oscillations and the phases entering $m_{\beta\beta}$ are separated at the level of the parametrization itself \cite{RodejohannValle2011,DingValle2025}. Ordinary oscillation data constrain only the combination in eq.~\eqref{eq:symmetricdelta}; they do not determine $\phi_{12}$ and $\phi_{13}$ separately. Since the magnitude-only MQLC ensemble is invariant under the phase reshuffling in eq.~\eqref{eq:psishift}, the present construction does not generate an additional prediction for $m_{\beta\beta}$ or for the Majorana phases without a further assumption fixing the complex structure of $\Psi$ and/or $V_c$.

The original symmetric formalism is also well suited to seesaw constructions with additional neutral fermions \cite{SchechterValle1980,RodejohannValle2011}. In the conventional high-scale seesaw limit the low-energy three-neutrino matrix is approximately unitary, which is the regime studied here. Low-scale seesaw realizations can instead induce observable non-unitarity in the effective $3\times3$ submatrix \cite{DingValle2025,AvilaEtAl2026}. Extending the present MQLC test to that case would require a different theoretical and statistical framework and is not assumed below.

\subsection{CKM representation}

The 2016 analysis used a Wolfenstein representation retaining terms through sixth order in $\lambda$ \cite{Wolfenstein:1983yz,Xing:1994is,Sharma:2016epj}. For the precision-era calculation we instead construct the CKM matrix from the exact standard parametrization with the all-order Wolfenstein definitions used by the Particle Data Group \cite{PDG2026},
\begin{equation}
 s_{12}^{q}=\lambda,
 \qquad
 s_{23}^{q}=A\lambda^2,
 \qquad
 s_{13}^{q}e^{-i\delta_q}=A\lambda^3(\rho-i\eta),
 \label{eq:wolfdef}
\end{equation}
where the barred and unbarred parameters are related by
\begin{equation}
 \rho+i\eta=
 \frac{\sqrt{1-A^2\lambda^4}\,(\bar\rho+i\bar\eta)}
 {\sqrt{1-\lambda^2\left[1-A^2\lambda^4(\bar\rho+i\bar\eta)\right]}}.
 \label{eq:rhobar}
\end{equation}
This construction is exactly unitary up to floating-point precision. The effect of replacing it by the historical sixth-order matrix is quantified in section~\ref{sec:robustness} and is numerically negligible for the present observables.

\subsection{What is and is not predicted}
\label{sec:predictionlogic}

Equation~\eqref{eq:qlc-master} may be inverted formally as
\begin{equation}
 U_{\rm PMNS}=(U_{\rm CKM}\Psi)^{-1}V_c.
 \label{eq:inverse}
\end{equation}
The historical work used a preferred $V_c$ texture to obtain constraints on less well measured lepton parameters \cite{Sharma:2016epj}. With present precision data, however, reconstructing $V_c$ from a PMNS sample and immediately applying eq.~\eqref{eq:inverse} to the same sample would return that PMNS matrix identically. We therefore separate the two tasks in this paper. The historical predictions are tested as published, whereas current data are used only to reconstruct and compare the $V_c$ texture.

A second point concerns averaging. If $V_c^{(k)}$ is unitary for every realization $k$, the matrix whose entries are $\langle|V_{c,ij}|\rangle$ need not itself be unitary. Consequently, we retain the complex matrix at every realization and perform all physical consistency checks sample by sample. Mean magnitude matrices are used only to display the centre of the ensemble.

\section{Historical benchmark and statistical procedure}
\label{sec:method}

\subsection{Historical benchmark}

The historical input set is taken from ref.~\cite{Sharma:2016epj}. It used
\begin{align}
 \lambda &=0.2255\pm0.0006, & A&=0.818\pm0.015,\nonumber\\
 \bar\rho&=0.124\pm0.024, & \bar\eta&=0.354\pm0.015,
 \label{eq:historicalckm}
\end{align}
and
\begin{align}
 \sin^2\theta_{13}&=0.0218\pm0.0010,\nonumber\\
 \sin^2\theta_{12}&=0.304^{+0.013}_{-0.012},\nonumber\\
 \sin^2\theta_{23}&=0.452^{+0.052}_{-0.028},\nonumber\\
 \delta&=306^{+39}_{-70}\;\text{degrees}.
 \label{eq:historicalpmns}
\end{align}
The mismatch phases were uniform over $[0,2\pi)$. We reconstruct this ensemble with split-normal representations of the quoted asymmetric one-sigma errors. To enforce exact sample-wise unitarity consistently in the historical and modern ensembles, the historical CKM parameters are inserted into the exact construction of eqs.~\eqref{eq:wolfdef}--\eqref{eq:rhobar}; the published sixth-order representation gives indistinguishable $|V_c|$ results at the precision relevant here.

The reconstructed mean magnitude matrix is
\begin{equation}
 \left\langle |V_c|\right\rangle_{2016}=\begin{pmatrix}
 0.8061 & 0.5407 & 0.1895\\
 0.4309 & 0.5889 & 0.6609\\
 0.3698 & 0.5828 & 0.7184
 \end{pmatrix}.
 \label{eq:historicalreconstructed}
\end{equation}
For comparison, the rounded weighted texture quoted in ref.~\cite{Sharma:2016epj} was
\begin{equation}
 |V_c|_{\rm pub}=\begin{pmatrix}
 0.80 & 0.54 & 0.18\\
 0.44 & 0.58 & 0.66\\
 0.36 & 0.58 & 0.72
 \end{pmatrix}.
 \label{eq:historicalpublished}
\end{equation}
The agreement is sufficient for the historical ensemble to serve as the reference distribution. We do not attempt to reproduce unpublished digits of the historical weighted entries by fitting them to the old atmospheric-angle result.

\subsection{Precision-era CKM and PMNS information}

The modern quark-sector inputs are the three-generation unitary global-fit parameters from the 2026 Review of Particle Physics \cite{PDG2026},
\begin{align}
 \lambda&=0.22517\pm0.00068, &
 A&=0.826^{+0.017}_{-0.015},\nonumber\\
 \bar\rho&=0.1576^{+0.0092}_{-0.0091}, &
 \bar\eta&=0.3556^{+0.0071}_{-0.0069}.
 \label{eq:pdg2026}
\end{align}
Because a covariance matrix for these four Wolfenstein parameters is not supplied with the quoted review fit, the uncertainties are sampled independently using split-normal distributions. The effect of fixing the CKM parameters to their central values is included among the robustness tests.

For the lepton sector we use the public NuFIT 6.1 numerical release \cite{Esteban:2024eli,NuFIT61}. The main analysis is the variant including the tabulated Super-Kamiokande and IceCube/DeepCore atmospheric likelihood information (the ``TByes'' release). The best-fit anchors of this solution are listed in table~\ref{tab:inputs}. Normal ordering (NO) and inverted ordering (IO) are treated separately throughout.

\begin{table}[t]
\centering
\caption{Historical lepton-sector inputs and precision-era best-fit anchors used in the analysis. The modern numerical sampling uses the NuFIT 6.1 projected likelihood tables rather than Gaussian errors around these best-fit values. CKM inputs are given in eqs.~\eqref{eq:historicalckm} and \eqref{eq:pdg2026}.}
\label{tab:inputs}
\begin{tabular}{lccc}
\toprule
Parameter & 2016 input & NuFIT 6.1 NO & NuFIT 6.1 IO\\
\midrule
$\sin^2\theta_{12}$ & $0.304^{+0.013}_{-0.012}$ & $0.3088$ & $0.3088$\\
$\sin^2\theta_{13}$ & $0.0218\pm0.0010$ & $0.02248$ & $0.02262$\\
$\sin^2\theta_{23}$ & $0.452^{+0.052}_{-0.028}$ & $0.470$ & $0.550$\\
$\delta$ [deg] & $306^{+39}_{-70}$ & $212$ & $274$\\
\bottomrule
\end{tabular}
\end{table}

\subsection{Projection-weighted ensemble}
\label{sec:projectionensemble}

NuFIT 6.1 provides one-, two- and three-dimensional $\Delta\chi^2$ projections, but not the complete four-dimensional numerical likelihood in $(\theta_{12},\theta_{13},\theta_{23},\delta)$ \cite{NuFIT61}. We therefore construct a factorized surrogate of the public marginalized information rather than claiming an exact reconstruction of the unreleased joint likelihood.

For each ordering, points in the released $(\sin^2\theta_{23},\delta)$ grid are sampled with probability proportional to
\begin{equation}
 w_{23,\delta}\propto \exp\!\left[-\frac{1}{2}\Delta\chi^2_{23,\delta}\right],
 \label{eq:weight23}
\end{equation}
and points in the released $(\sin^2\theta_{13},\sin^2\theta_{12})$ grid are sampled analogously. The two normalized marginalized distributions are then combined under a factorization approximation. This preserves the dominant non-Gaussian atmospheric-angle--CP correlation and the published solar--reactor correlation without interpreting the two projected $\chi^2$ surfaces as independent experimental likelihoods. The approximation is tested in section~\ref{sec:robustness} by replacing the second two-dimensional projection with independent one-dimensional projections and by fixing the well-measured solar and reactor angles.

For every draw we generate $U_{\rm CKM}$, $U_{\rm PMNS}$ and the mismatch phases and store the complete complex matrix
\begin{equation}
 V_c^{(k)}=U_{\rm CKM}^{(k)}\Psi^{(k)}U_{\rm PMNS}^{(k)}.
 \label{eq:samplevc}
\end{equation}
The primary ensembles contain $3\times10^5$ realizations for each mass ordering and the same number for the historical reconstruction. The unitarity diagnostic
\begin{equation}
 \epsilon_U^{(k)}=\left\|V_c^{(k)}V_c^{(k)\dagger}-\mathbb{1}\right\|_F
 \label{eq:unitarityresidual}
\end{equation}
remains below $1.1\times10^{-15}$ in the sampled checks, confirming unitary matrix construction to floating-point precision.

We use several descriptive diagnostics. For two distributions $p_A(x)$ and $p_B(x)$ of a given modulus, the overlap coefficient is
\begin{equation}
 \mathcal O_{ij}(A,B)=\int_0^1
 \min\!\left[p_A(|V_{c,ij}|),p_B(|V_{c,ij}|)\right]
 \,d|V_{c,ij}|,
 \label{eq:overlap}
\end{equation}
with $0\leq\mathcal O_{ij}\leq1$. We also define the normalized central-texture drift
\begin{equation}
 S_O=\frac{\left\|\langle|V_c|\rangle_O-
 \langle|V_c|\rangle_{2016}\right\|_F}
 {\left\|\langle|V_c|\rangle_{2016}\right\|_F},
 \label{eq:texturedrift}
\end{equation}
where $O$ denotes NO or IO. Equation~\eqref{eq:texturedrift} is a descriptive measure of matrix change and is not interpreted as a statistical significance.

\section{Retrospective test of historical MQLC predictions}
\label{sec:historical-test}

\subsection{The narrow 2016 atmospheric-angle prediction}

The cleanest prospective test uses the one-dimensional NuFIT 6.1 $\Delta\chi^2(\sin^2\theta_{23})$ profiles directly and is independent of the factorization approximation described above. Each ordering is referenced to its own local minimum. Evaluating the profile at the historical central prediction in eq.~\eqref{eq:oldtheta23} gives
\begin{equation}
 \Delta\chi^2_{\rm NO}=17.99,
 \qquad
 \Delta\chi^2_{\rm IO}=20.43.
 \label{eq:oldpenalties}
\end{equation}
Under the usual one-parameter Wilks interpretation, $\sqrt{\Delta\chi^2}$ corresponds approximately to $4.24\sigma$ and $4.52\sigma$, respectively. We quote the $\Delta\chi^2$ values as the primary result because the conversion to a Gaussian-equivalent significance is only approximate.

Even the upper edge of the historical one-sigma interval, $\sin^2\theta_{23}=0.4267$, has
\begin{equation}
 \Delta\chi^2_{\rm NO}=15.13,
 \qquad
 \Delta\chi^2_{\rm IO}=17.56.
 \label{eq:oldupper}
\end{equation}
Thus the narrow lower-octant prediction is strongly disfavoured by the present global fit. This conclusion is not generated only by the additional tabulated atmospheric information: with the NuFIT 6.1 ``TBoff'' release, the central historical value still gives $\Delta\chi^2=14.78$ for NO and $17.52$ for IO.

\begin{figure}[t]
 \centering
 \includegraphics[width=0.86\textwidth]{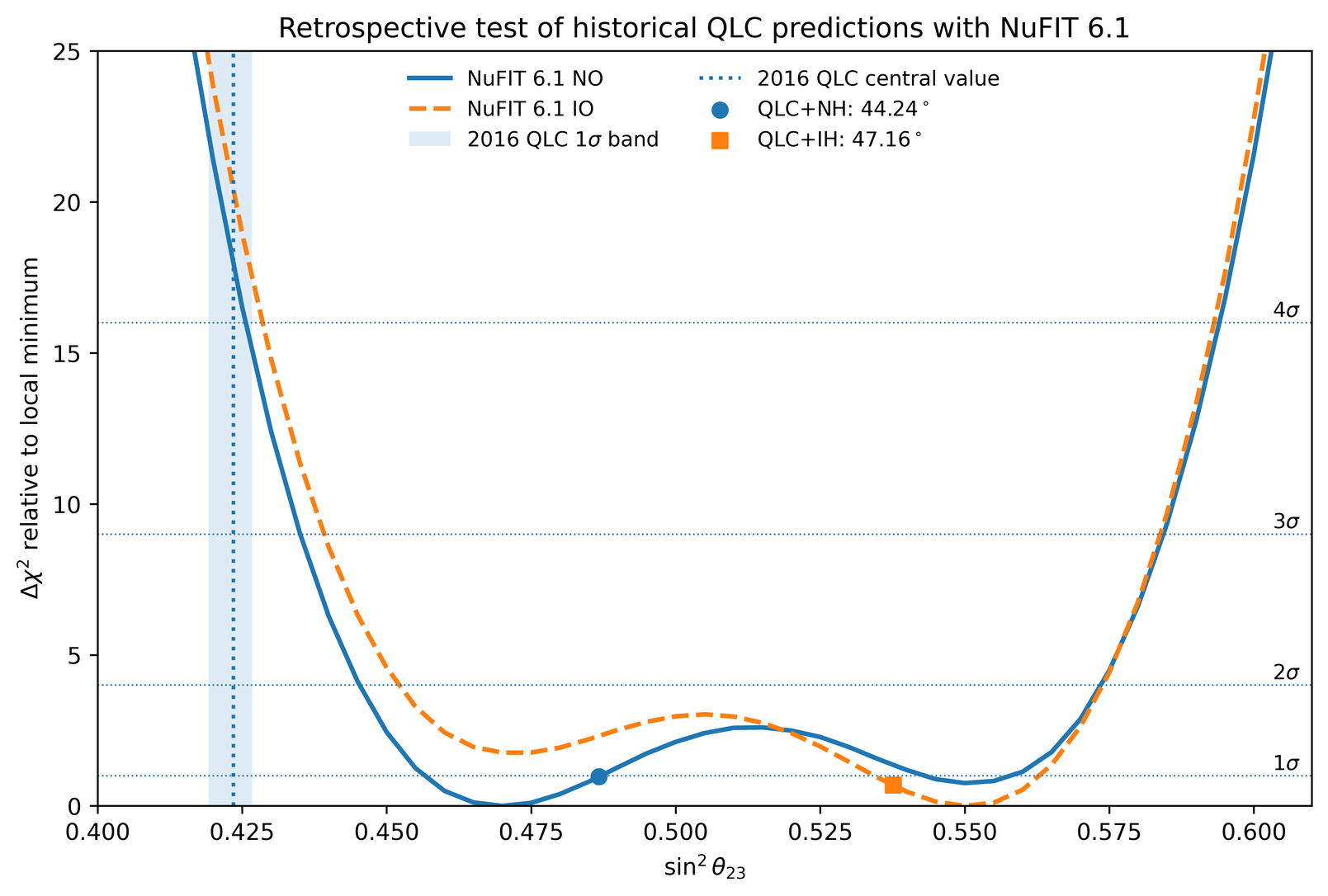}
 \caption{Retrospective test of historical QLC atmospheric-angle predictions using the NuFIT 6.1 one-dimensional $\Delta\chi^2$ profiles. The curves are measured relative to the local minimum of each ordering. The shaded band denotes the 2016 one-sigma QLC interval, while the markers show the subsequent ordering-dependent predictions. The original narrow lower-octant result is strongly disfavoured, whereas the ordering-dependent central values remain close to the current preferred regions.}
 \label{fig:theta23test}
\end{figure}

Figure~\ref{fig:theta23test} shows the likelihood profiles together with the historical one-sigma band. Table~\ref{tab:retrospective} summarizes the principal numerical tests.

\begin{table}[t]
\centering
\caption{Retrospective comparison of published QLC atmospheric-angle predictions with the NuFIT 6.1 IC24+SK-atm one-dimensional profile. Gaussian-equivalent values are shown only as the approximate one-parameter quantity $\sqrt{\Delta\chi^2}$.}
\label{tab:retrospective}
\begin{tabular}{lccc}
\toprule
Prediction tested & Ordering & $\Delta\chi^2$ & $\sqrt{\Delta\chi^2}$\\
\midrule
2016 central value, $\sin^2\theta_{23}=0.4235$ & NO & 17.99 & 4.24\\
2016 central value, $\sin^2\theta_{23}=0.4235$ & IO & 20.43 & 4.52\\
2016 upper one-sigma edge, $0.4267$ & NO & 15.13 & 3.89\\
2016 upper one-sigma edge, $0.4267$ & IO & 17.56 & 4.19\\
Ordering-aware $\theta_{23}=44.24^\circ$ & NO & 0.97 & 0.98\\
Ordering-aware $\theta_{23}=47.16^\circ$ & IO & 0.68 & 0.83\\
\bottomrule
\end{tabular}
\end{table}

\subsection{Ordering-dependent follow-up prediction}

The later QLC analysis that separated the two hierarchy hypotheses gave central values $44.24^\circ$ and $47.16^\circ$ \cite{Sharma:2018hierarchy}. They correspond to $\sin^2\theta_{23}=0.48674$ and $0.53766$. Evaluating the current profiles gives
\begin{equation}
 \Delta\chi^2_{\rm NO}(44.24^\circ)=0.97,
 \qquad
 \Delta\chi^2_{\rm IO}(47.16^\circ)=0.68.
 \label{eq:hierarchytest}
\end{equation}
These values remain well within the current preferred regions. The contrast with eq.~\eqref{eq:oldpenalties} is useful: present precision data discriminate strongly between the narrow 2016 realization and the later ordering-dependent realization rather than providing a uniform yes-or-no verdict on all QLC implementations.

\subsection{CP invariants and LNV phase directions}
\label{sec:cpresult}

We first propagate the projection-weighted PMNS ensemble to the rephasing-invariant Jarlskog quantity \cite{Jarlskog:1985ht},
\begin{equation}
 J_{\rm CP}=c_{12}s_{12}c_{23}s_{23}c_{13}^2s_{13}\sin\delta.
 \label{eq:jcp}
\end{equation}
In the symmetric parametrization this is equivalently given by eq.~\eqref{eq:symmetricJ}. The median values remain
\begin{equation}
 J_{\rm CP}^{\rm med}=-0.0143\quad({\rm NO}),
 \qquad
 J_{\rm CP}^{\rm med}=-0.0326\quad({\rm IO}).
 \label{eq:jmedian}
\end{equation}
Within the derived ensembles, $90.5\%$ of NO samples and $33.3\%$ of IO samples satisfy the historical condition $|J_{\rm CP}|<0.0315$. These fractions are diagnostics of the projection-weighted surrogate and should not be confused with direct NuFIT confidence levels.

The historical 2016 study also quoted
\begin{equation}
 |S_1|<0.12,\qquad |S_2|<0.08,
 \label{eq:oldS12}
\end{equation}
where, in its phase convention,
\begin{align}
 S_1&=\frac{1}{2}\cos\theta_{12}\sin2\theta_{13}\sin(\phi_1+\delta),\\
 S_2&=\frac{1}{2}\sin\theta_{12}\sin2\theta_{13}\sin(\phi_2+\delta).
 \label{eq:S12definitions}
\end{align}
Unlike $J_{\rm CP}$, these quantities depend on Majorana-phase directions that are not measured by ordinary oscillations. If those phases are allowed to vary freely, the absolute kinematic ceilings are simply
\begin{align}
 |S_1|_{\rm max}&=\frac{1}{2}\cos\theta_{12}\sin2\theta_{13},\\
 |S_2|_{\rm max}&=\frac{1}{2}\sin\theta_{12}\sin2\theta_{13}.
 \label{eq:S12ceilings}
\end{align}
At the central inputs used in the 2016 analysis,
\begin{equation}
 |S_1|_{\rm max}=0.12183,\qquad |S_2|_{\rm max}=0.08052,
 \label{eq:S12oldceilings}
\end{equation}
so the reported limits $0.12$ and $0.08$ correspond to about $98.5\%$ and $99.4\%$ of the respective phase-agnostic ceilings. With the NuFIT 6.1 best-fit angles the ceilings are
\begin{align}
 {\rm NO:}\quad& |S_1|_{\rm max}=0.12324,\qquad |S_2|_{\rm max}=0.08238,\\
 {\rm IO:}\quad& |S_1|_{\rm max}=0.12362,\qquad |S_2|_{\rm max}=0.08263.
 \label{eq:S12modernceilings}
\end{align}
The historical $S_1$ and $S_2$ limits therefore lie very close to the maximum ranges already allowed by the mixing-angle prefactors when the Majorana phases are unrestricted. They should not be interpreted as precision-era constraints on Majorana CP violation. A meaningful present-day restriction on the LNV phase directions would require either LNV experimental information or an additional model assumption that fixes part of the complex MQLC phase structure.

\section{Precision-era reconstruction of the MQLC correlation matrix}
\label{sec:vc-results}

\subsection{Element distributions}

\begin{figure}[t]
 \centering
 \includegraphics[width=\textwidth]{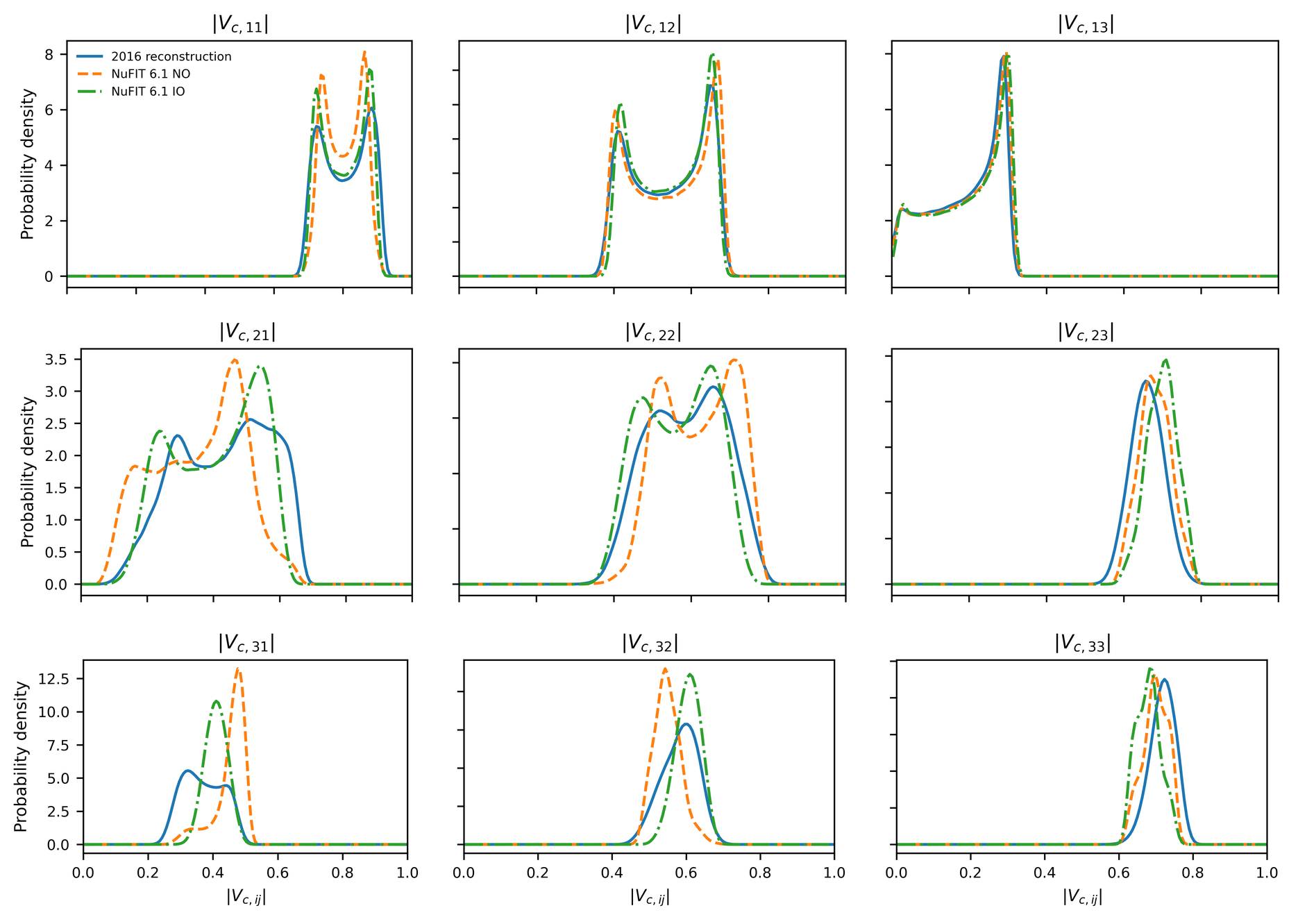}
 \caption{Probability-density distributions of the nine $|V_{c,ij}|$ elements for the historical reconstruction and the NuFIT 6.1 projection-weighted NO and IO ensembles. The first row is comparatively stable, while the larger shifts occur in the lower two rows.}
 \label{fig:vcpdfs}
\end{figure}

The likelihood-weighted mean magnitude matrices are
\begin{equation}
\left\langle|V_c|\right\rangle_{\rm NO}=
\begin{pmatrix}
0.8026&0.5449&0.1936\\
0.3560&0.6174&0.6807\\
0.4532&0.5496&0.6984
\end{pmatrix},
\label{eq:vcno}
\end{equation}
and
\begin{equation}
\left\langle|V_c|\right\rangle_{\rm IO}=
\begin{pmatrix}
0.8029&0.5436&0.1970\\
0.4059&0.5658&0.6982\\
0.4099&0.6051&0.6800
\end{pmatrix}.
\label{eq:vcio}
\end{equation}
Table~\ref{tab:vcintervals} gives the central 68\% intervals. Figure~\ref{fig:vcpdfs} compares the complete historical and precision-era one-dimensional distributions.

\begin{table}[t]
\centering
\caption{Precision-era projection-weighted magnitude distributions. Entries show the ensemble mean followed by the central 68\% interval. The displayed mean matrix is a descriptive summary; it is not itself interpreted as a unitary matrix.}
\label{tab:vcintervals}
\begin{tabular}{ccc}
\toprule
Element & NO & IO\\
\midrule
$|V_{c,11}|$ & $0.8026\ [0.7386,0.8646]$ & $0.8029\ [0.7263,0.8773]$\\
$|V_{c,12}|$ & $0.5449\ [0.4184,0.6615]$ & $0.5436\ [0.4281,0.6501]$\\
$|V_{c,13}|$ & $0.1936\ [0.0762,0.2926]$ & $0.1970\ [0.0777,0.2970]$\\
$|V_{c,21}|$ & $0.3560\ [0.1919,0.4921]$ & $0.4059\ [0.2443,0.5501]$\\
$|V_{c,22}|$ & $0.6174\ [0.5112,0.7224]$ & $0.5658\ [0.4582,0.6691]$\\
$|V_{c,23}|$ & $0.6807\ [0.6391,0.7220]$ & $0.6982\ [0.6584,0.7383]$\\
$|V_{c,31}|$ & $0.4532\ [0.4152,0.4924]$ & $0.4099\ [0.3747,0.4453]$\\
$|V_{c,32}|$ & $0.5496\ [0.5131,0.5844]$ & $0.6051\ [0.5706,0.6395]$\\
$|V_{c,33}|$ & $0.6984\ [0.6620,0.7353]$ & $0.6800\ [0.6442,0.7131]$\\
\bottomrule
\end{tabular}
\end{table}

Several features are visible directly. First, the first row changes very little in its central texture. In particular, the non-zero $|V_{c,13}|$ found after the reactor-angle measurement remains a persistent feature. Second, the principal evolution occurs in the lower block. In NO, $|V_{c,21}|$ moves downward while $|V_{c,31}|$ moves upward relative to the reconstructed historical distribution. In IO, the rearrangement is more distributed among $|V_{c,23}|$, $|V_{c,31}|$ and $|V_{c,33}|$.

\subsection{Mean-texture drift}

\begin{figure}[t]
 \centering
 \includegraphics[width=0.92\textwidth]{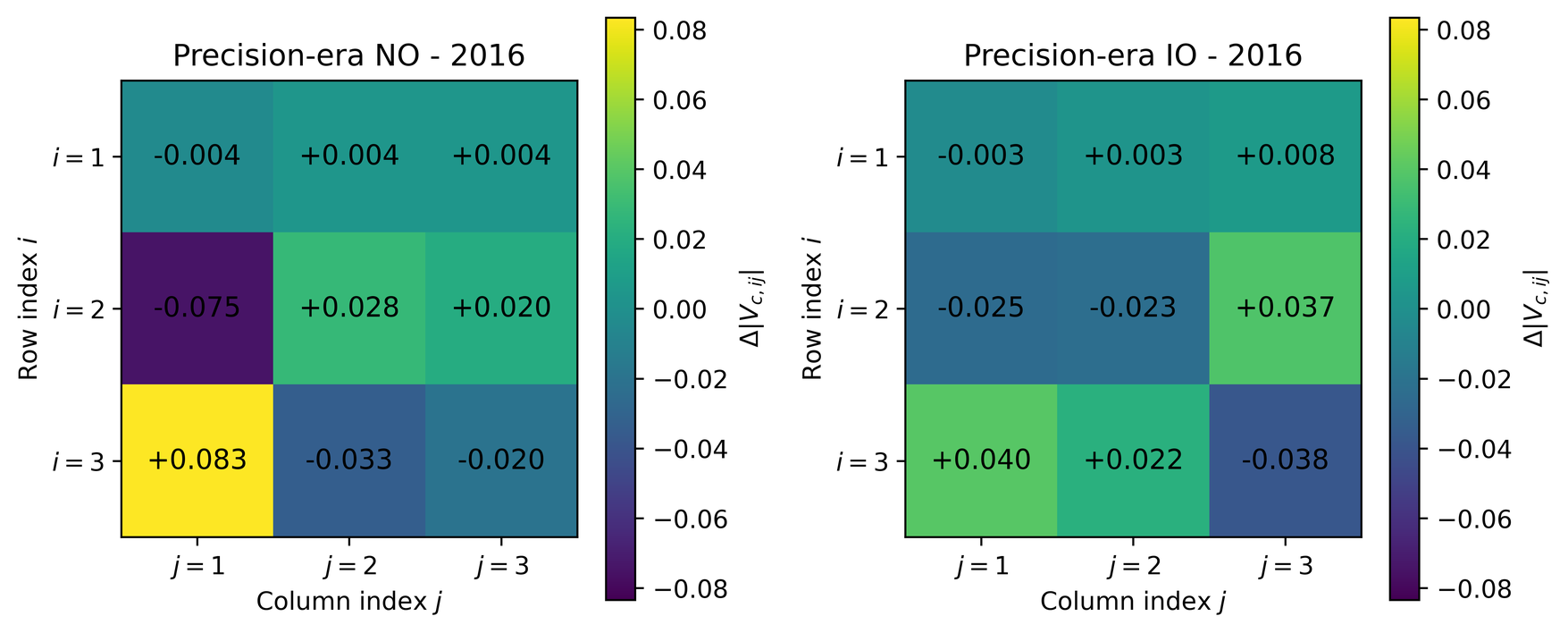}
 \caption{Signed element-by-element change in the mean magnitude texture relative to the reconstructed 2016 ensemble. The precision-era shifts are strongly localized in the second and third rows.}
 \label{fig:shift}
\end{figure}

The signed changes in the central magnitude texture are shown in figure~\ref{fig:shift}. The normalized drift of eq.~\eqref{eq:texturedrift} is
\begin{equation}
 S_{\rm NO}=7.22\%,
 \qquad
 S_{\rm IO}=4.59\%.
 \label{eq:drifts}
\end{equation}
The direct NO--IO separation, normalized to the historical texture norm, is $6.05\%$.

The small overall numbers hide a highly non-uniform redistribution. For NO, rows 1, 2 and 3 contribute $0.31\%$, $44.47\%$ and $55.22\%$, respectively, to the squared mean-texture change. For IO the corresponding contributions are $1.20\%$, $41.10\%$ and $57.70\%$. Hence about 99\% of the mean-texture evolution is carried by the lower two rows in either ordering.

\subsection{Distribution overlap and ordering dependence}

\begin{figure}[t]
 \centering
 \includegraphics[width=\textwidth]{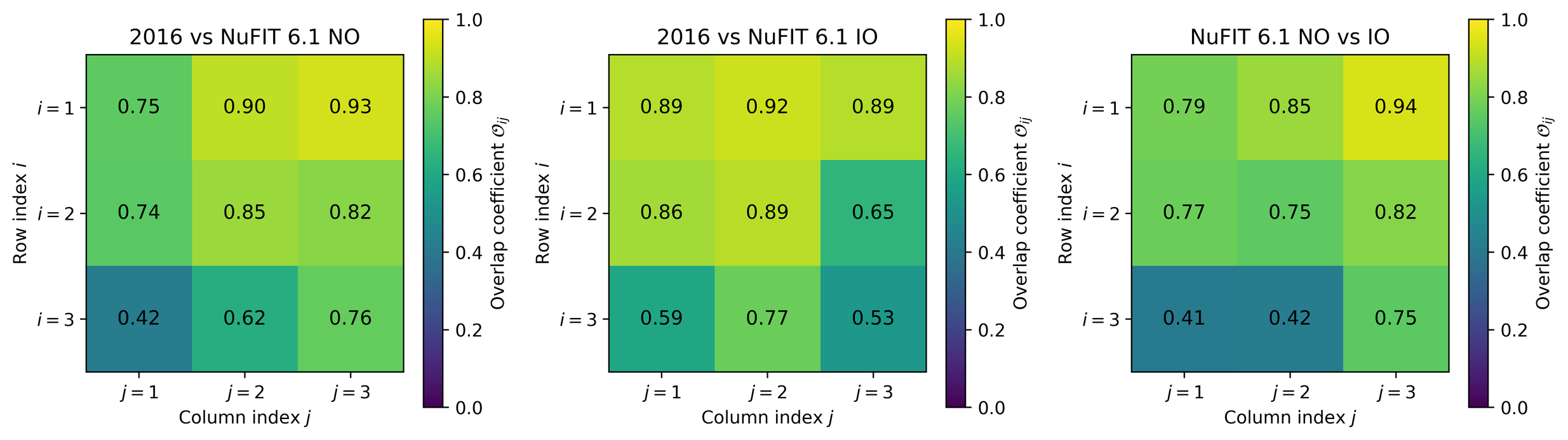}
 \caption{Distribution-overlap coefficients for the historical-to-NO, historical-to-IO and direct NO-to-IO comparisons. Values close to unity indicate strongly overlapping distributions. The third row contains the most pronounced historical migration and the strongest present ordering dependence.}
 \label{fig:overlap}
\end{figure}

A shift of the mean does not by itself establish that two distributions are distinct. Figure~\ref{fig:overlap} therefore displays the overlap coefficients of eq.~\eqref{eq:overlap}. For the historical-to-NO comparison, the smallest overlap is
\begin{equation}
 \mathcal O_{31}(2016,{\rm NO})=0.422,
 \label{eq:histnooverlap}
\end{equation}
followed by $\mathcal O_{32}=0.623$. For the historical-to-IO comparison, the largest migration occurs in $V_{c,33}$, with
\begin{equation}
 \mathcal O_{33}(2016,{\rm IO})=0.535,
 \label{eq:histiooverlap}
\end{equation}
followed by $V_{c,31}$ and $V_{c,23}$.

The direct ordering comparison gives
\begin{equation}
 \mathcal O_{31}({\rm NO,IO})=0.409,
 \qquad
 \mathcal O_{32}({\rm NO,IO})=0.420,
 \label{eq:noiooverlap}
\end{equation}
which are the two smallest element-wise overlaps and identify the third row as the most ordering-sensitive sector of the reconstructed texture. The shifts are in opposite directions:
\begin{equation}
 \langle|V_{c,31}|\rangle_{\rm NO}>
 \langle|V_{c,31}|\rangle_{\rm IO},
 \qquad
 \langle|V_{c,32}|\rangle_{\rm IO}>
 \langle|V_{c,32}|\rangle_{\rm NO}.
 \label{eq:orderingdirections}
\end{equation}
These elements should be described as ordering-sensitive \emph{within the reconstructed QLC texture}; they do not constitute an independent mass-ordering observable because the reconstruction already uses ordering-dependent PMNS likelihoods.

\section{TBM/BM comparison and robustness}
\label{sec:tbm-robust}

\subsection{Sample-wise proximity to reference textures}

The historical MQLC analysis described $V_c$ as closer to TBM than BM \cite{Sharma:2016epj,Shimizu:2010qy}. To test whether this statement remains meaningful at the distribution level, we use the reference magnitude matrices
\begin{equation}
 |V_{\rm TBM}|=\begin{pmatrix}
 \sqrt{2/3}&1/\sqrt3&0\\
 1/\sqrt6&1/\sqrt3&1/\sqrt2\\
 1/\sqrt6&1/\sqrt3&1/\sqrt2
 \end{pmatrix},
 \qquad
 |V_{\rm BM}|=\begin{pmatrix}
 1/\sqrt2&1/\sqrt2&0\\
 1/2&1/2&1/\sqrt2\\
 1/2&1/2&1/\sqrt2
 \end{pmatrix}.
 \label{eq:tbmbm}
\end{equation}
For every physical sample we calculate
\begin{equation}
 D_X^{(k)}=\left\||V_c^{(k)}|-|V_X|\right\|_F,
 \qquad X\in\{{\rm TBM,BM}\},
 \label{eq:distances}
\end{equation}
and define $\Delta D=D_{\rm BM}-D_{\rm TBM}$. Positive $\Delta D$ corresponds to a sample closer to TBM.

\begin{figure}[t]
 \centering
 \includegraphics[width=0.80\textwidth]{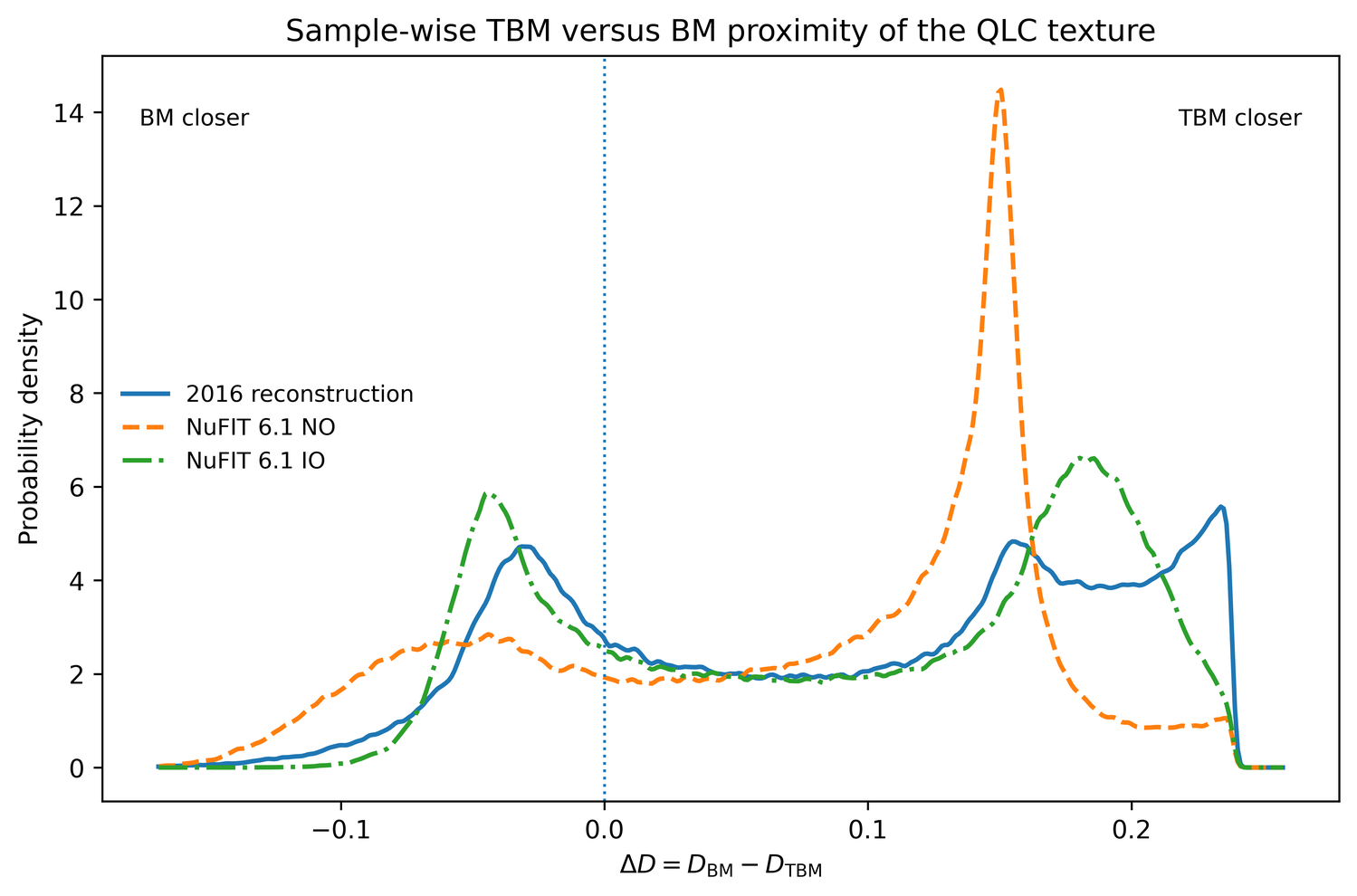}
 \caption{Sample-wise TBM-versus-BM proximity. Positive $\Delta D=D_{\rm BM}-D_{\rm TBM}$ indicates a $V_c$ realization closer to the TBM reference texture. Approximately 72\% of both precision-era ensembles fall on the TBM side, very close to the reconstructed historical fraction.}
 \label{fig:tbmbm}
\end{figure}

The fractions satisfying $D_{\rm TBM}<D_{\rm BM}$ are
\begin{equation}
 73.21\%\quad(2016),
 \qquad
 71.95\%\quad({\rm NO}),
 \qquad
 72.17\%\quad({\rm IO}).
 \label{eq:tbmprobability}
\end{equation}
The distributions are shown in figure~\ref{fig:tbmbm}. The result supports a persistent but moderate TBM-over-BM tendency; it should not be read as evidence that the correlation matrix is exactly TBM.

\subsection{Robustness tests}
\label{sec:robustness}

Table~\ref{tab:robustness} summarizes numerical tests of the main analysis choices. Replacing the $(\theta_{13},\theta_{12})$ two-dimensional projection by independently sampled one-dimensional NuFIT profiles changes any mean $|V_{c,ij}|$ by at most $3.3\times10^{-4}$ in NO and $5.9\times10^{-4}$ in IO. Fixing the same two precisely measured angles to their best-fit values produces shifts below $1.2\times10^{-3}$. This confirms that the texture evolution is driven mainly by the atmospheric-angle--CP structure rather than by residual treatment of the solar and reactor angles.

Sampling the quoted one-dimensional PDG CKM errors rather than fixing the CKM parameters changes mean moduli by less than $9.2\times10^{-4}$. More importantly, replacing the exact CKM construction by the historical $O(\lambda^6)$ form changes the mean $|V_c|$ entries by only $1.3\times10^{-5}$. The truncated CKM matrix has a unitarity residual of about $4.7\times10^{-5}$ at the current central point, compared with $\sim10^{-16}$ for the exact construction, so the exact representation is formally preferable even though the phenomenological difference is negligible.

The atmospheric-data choice has the largest, though still modest, effect among the tested variations. Replacing TByes by TBoff changes the mean entries by at most $0.0054$ in NO and $0.0095$ in IO. The smallest element-wise TByes--TBoff overlap is $0.933$ for NO and $0.880$ for IO. The central conclusion about the historical prediction is also unchanged, as already noted in section~\ref{sec:historical-test}.

Finally, convergence was assessed with prefixes of the same $3\times10^5$ ensembles. At $2\times10^5$ samples the maximum change of any mean entry relative to the full ensemble is $7.5\times10^{-5}$ for NO and $2.3\times10^{-4}$ for IO.

\begin{table}[t]
\centering
\caption{Selected robustness diagnostics. ``Max. shift'' denotes the maximum absolute change among the nine mean $|V_{c,ij}|$ entries.}
\label{tab:robustness}
\begin{tabular}{lcc}
\toprule
Test & NO & IO\\
\midrule
2D $(\theta_{13},\theta_{12})$ vs independent 1D profiles, max. shift
& $3.3\times10^{-4}$ & $5.9\times10^{-4}$\\
2D projection vs fixed $\theta_{12},\theta_{13}$, max. shift
& $1.2\times10^{-3}$ & $1.1\times10^{-3}$\\
Sampled vs fixed CKM inputs, max. shift
& $9.1\times10^{-4}$ & $8.7\times10^{-4}$\\
TByes vs TBoff, max. shift
& $5.4\times10^{-3}$ & $9.5\times10^{-3}$\\
TByes vs TBoff, minimum overlap
& $0.933$ & $0.880$\\
Exact CKM vs $O(\lambda^6)$, max. shift
& $1.3\times10^{-5}$ & $1.3\times10^{-5}$\\
200k vs 300k samples, max. mean deviation
& $7.5\times10^{-5}$ & $2.3\times10^{-4}$\\
\bottomrule
\end{tabular}
\end{table}

\section{Discussion}
\label{sec:discussion}

The precision-era update leads to a more differentiated picture of the modified QLC construction than either a simple confirmation or rejection. The narrow 2016 lower-octant prediction is now strongly disfavoured by the direct NuFIT 6.1 atmospheric-angle likelihood, while the later ordering-dependent central values remain close to the current minima. This contrast indicates that the most restrictive atmospheric-angle result was not a robust consequence of the broad QLC idea alone; it depended on the particular historical texture extraction and prediction procedure.

At the same time, the empirical $V_c$ texture changes much less dramatically. The first row remains especially persistent, the overall central-texture drift is at the few-to-seven-percent level, and the historical preference for TBM over BM survives almost unchanged when assessed sample by sample. The main evolution is instead localized in the lower two rows, which contain roughly 99\% of the squared mean-texture change. This localization is physically understandable in the limited sense that these entries inherit the present uncertainty and ordering dependence of the atmospheric and CP sector. It should not be interpreted as a new standalone probe of the ordering.

The treatment of the correlation matrix itself is also important. A matrix formed from nine separately averaged moduli is generally not unitary, even though every underlying complex $V_c$ is. The present analysis therefore avoids using such an averaged magnitude matrix as the input of eq.~\eqref{eq:inverse}. This prevents the current PMNS data from being fed into $V_c$ and then returned as an apparent ``prediction.'' The old atmospheric-angle value is instead tested prospectively, which is statistically and conceptually cleaner.

The symmetric parametrization sharpens the meaning of the phase-convention issue. The magnitude texture reconstructed here is not merely numerically insensitive to Majorana phases; with unrestricted mismatch phases it is exactly covariant under the redistribution of lepton phases between the symmetric mixing matrix and the diagonal mismatch matrix. Thus the persistence of $|V_c|$ is a statement about the correlation of mixing magnitudes, not a hidden prediction of the LNV phase sector. The contrast between eqs.~\eqref{eq:oscphasecancel} and \eqref{eq:lnvamplitude} makes this physical separation explicit: oscillation observables discard the column phases, whereas LNV amplitudes retain them. If a future ultraviolet realization fixes $\Psi$ or specifies a complex $V_c$ rather than only its modulus structure, the phase absorption in eq.~\eqref{eq:Vcsym} would no longer be freely available and genuinely new Majorana-phase correlations could emerge.

The same distinction clarifies the historical $S_1$ and $S_2$ limits. Their proximity to the unrestricted-phase ceilings in eqs.~\eqref{eq:S12oldceilings} and \eqref{eq:S12modernceilings} shows that they do not constitute strong present-day information on the Majorana phases. The symmetric representation makes explicit why oscillation data can update the Dirac combination in eq.~\eqref{eq:symmetricdelta} while leaving two independent LNV phase directions undetermined. Neutrinoless double-beta decay then provides the cleanest illustration: $m_{\beta\beta}$ depends directly on $\phi_{12}$ and $\phi_{13}$, but the phase-covariant magnitude-only MQLC construction supplies no additional restriction on either. Consequently, a generic $m_{\beta\beta}$ band obtained by scanning unrestricted phases would be the standard three-neutrino Majorana band, not an MQLC prediction. A narrower MQLC-specific LNV region would require additional phase-fixing information from a specified flavour/seesaw realization or from independent LNV data.

There are several limitations. First, the NuFIT public release provides projected likelihoods rather than the complete multidimensional likelihood. Our factorized surrogate preserves the strongest released two-dimensional structures, but cannot reproduce correlations between the $(\theta_{23},\delta)$ and $(\theta_{13},\theta_{12})$ pairs. The small changes found under alternative treatments of the well-measured solar and reactor parameters suggest that this limitation does not drive the reported texture results, but a future public higher-dimensional likelihood would allow a cleaner joint reconstruction.

Second, the analysis remains within a unitary three-neutrino description. The Schechter--Valle formalism naturally accommodates additional neutral fermions, and both high- and low-scale seesaw settings provide broader contexts in which the phase structure can be studied \cite{SchechterValle1980,DingValle2025,AvilaEtAl2026}. In a low-scale seesaw, effective non-unitarity of the light-neutrino submatrix could modify the correlation relation itself; treating that possibility would require a separate extension rather than a reinterpretation of the present ensemble.

Third, all comparisons are performed with low-energy mixing parameters. If QLC is imposed at a unification scale, renormalization-group evolution can modify the relation, with a model dependence that grows for particular mass spectra and high-scale parameter choices \cite{Kang:2005ic,Antusch:2005gp}. Incorporating such running would answer a different question from the retrospective one addressed here and is left outside the present scope.

Within these limitations, the decade-long comparison remains informative. The underlying magnitude texture has proved more durable than the narrow atmospheric-angle prediction extracted from it, while the LNV phase sector remains genuinely open. This three-way separation is the main phenomenological message of the revised analysis.

\section{Conclusions}
\label{sec:conclusions}

We have reassessed the three-generation modified QLC construction using the non-trivial correlation relation of eq.~\eqref{eq:qlc-master}, current PDG CKM inputs and the public NuFIT 6.1 likelihood release. The analysis was deliberately separated into a retrospective prediction test, a precision-era magnitude-texture reconstruction and an explicit examination of the lepton-phase convention.

The published 2016 central prediction $\sin^2\theta_{23}=0.4235$ receives NuFIT 6.1 penalties of $17.99$ in NO and $20.43$ in IO and is therefore strongly disfavoured. This conclusion remains when the additional tabulated SK and IC24 atmospheric likelihoods are removed. By contrast, the later ordering-dependent QLC central values of $44.24^\circ$ and $47.16^\circ$ have penalties of only $0.97$ and $0.68$ in their corresponding orderings.

The modern correlation-matrix reconstruction gives a different message. Its first row remains close to the historical texture, while approximately 99\% of the squared mean-texture evolution is concentrated in the lower two rows. The normalized central-texture drift is $7.22\%$ for NO and $4.59\%$ for IO. The strongest direct ordering dependence occurs in $V_{c,31}$ and $V_{c,32}$, whose NO--IO overlap coefficients are $0.409$ and $0.420$. The historical TBM-over-BM tendency also persists: about 72\% of present samples are closer to the TBM reference texture for either ordering.

Using the symmetric Schechter--Valle parametrization, we have additionally shown that the ordinary oscillation phase is the invariant combination $\phi_{13}-\phi_{12}-\phi_{23}$, whereas two independent phase directions remain relevant to LNV amplitudes. The advantage is transparent at amplitude level: Majorana column phases cancel from oscillation products but survive in $U_{\alpha i}U_{\beta i}$ combinations entering LNV processes, with neutrinoless double-beta decay depending directly on $\phi_{12}$ and $\phi_{13}$ through $m_{\beta\beta}$. Because the MQLC construction used here treats the diagonal mismatch phases as unrestricted, the left rephasings that connect the symmetric and PDG representations are absorbed exactly into $\Psi$, while the remaining right rephasings only change complete columns of $V_c$. The full ensemble of $|V_{c,ij}|$ is therefore invariant under the change of parametrization. This analytic result makes explicit which part of the reconstructed texture is convention robust and which phase information is absent from the present framework.

The same separation clarifies the historical Majorana-sensitive invariants. Their quoted limits, $|S_1|<0.12$ and $|S_2|<0.08$, lie close to the kinematic maxima implied by the measured $\theta_{12}$ and $\theta_{13}$ when the Majorana phases are unrestricted. Present oscillation data therefore do not turn these quantities into precision constraints on Majorana CP violation. Any predictive connection between MQLC and neutrinoless double-beta decay would require additional information that fixes the complex phase structure, for example through a specified flavour/seesaw realization or direct LNV input.

Taken together, the results distinguish three logically separate statements: the narrow 2016 atmospheric-angle prediction is strongly disfavoured; the broader magnitude texture remains comparatively stable and predominantly TBM-like; and the Majorana/LNV phase sector is not fixed by the magnitude-only MQLC construction. This separation provides a more precise statement of what survives, what fails and what remains genuinely undetermined in modified quark--lepton complementarity.

\appendix

\section{Numerical definitions and reproducibility}
\label{app:numerics}

The primary reported ensembles use $3\times10^5$ samples for each ordering. For a released grid point $g$ with projected value $\Delta\chi_g^2$, sampling is performed from the normalized discrete density
\begin{equation}
 p_g=\frac{\exp(-\Delta\chi_g^2/2)}{\sum_h\exp(-\Delta\chi_h^2/2)}.
 \label{eq:discreteweights}
\end{equation}
The overlap coefficients in eq.~\eqref{eq:overlap} are evaluated numerically from common histograms spanning $0\leq|V_{c,ij}|\leq1$ with 500 bins. Central 68\% intervals correspond to the 16th and 84th percentiles. Random-number seeds are fixed in the supplied analysis script.

The retrospective atmospheric-angle tests do not rely on Monte Carlo sampling: the published historical values are evaluated by linear interpolation of the released one-dimensional NuFIT 6.1 $\Delta\chi^2(\sin^2\theta_{23})$ grid after subtracting the local minimum for the corresponding ordering.

\section*{Data and code availability}

The NuFIT 6.1 numerical $\Delta\chi^2$ tables used in this work are publicly available from the NuFIT project \cite{NuFIT61}; the CKM inputs are taken from the 2026 Review of Particle Physics \cite{PDG2026}. No new experimental data were generated. The analysis scripts that parse the released NuFIT tables and reproduce the derived ensembles, diagnostic quantities and figures are supplied as supplementary material with this submission.

\end{document}